\documentclass[aps,prb,twocolumn,superscriptaddress,unsortedaddress,
preprintnumbers,showpacs]{revtex4}

\usepackage{graphicx}% Include figure files

\expandafter\let\expandafter\savedselectfont\csname selectfont
\endcsname
\expandafter\def\csname selectfont \endcsname{%
   \savedselectfont
   \hyphenchar\font=-1\relax}

\begin{document}

\title{First-principles study on thermodynamical stability of metal 
borohydrides:\\ Aluminum borohydride ${\rm Al(BH_4)_3}$}

\author{Kazutoshi Miwa}
%\email{miwa@cmp.tytlabs.co.jp}
\affiliation{
Toyota Central Research $\&$ Development Laboratories, Inc.,
Nagakute, Aichi 480-1192, Japan}

\author{Nobuko Ohba}
\affiliation{
Toyota Central Research $\&$ Development Laboratories, Inc.,
Nagakute, Aichi 480-1192, Japan}

\author{Yuko Nakamori}
\affiliation{
Institute for Materials Research, Tohoku University,
Sendai 980-8577, Japan}

\author{Shin-ichi Towata}
\affiliation{
Toyota Central Research $\&$ Development Laboratories, Inc.,
Nagakute, Aichi 480-1192, Japan}

\author{Andreas Z\"{u}ttel}
\affiliation{
Physics Department, University of Fribourg, Perolles, Switzerland}
\affiliation{
Department of Environment, Energy and Mobility, EMPA, D\"{u}bendorf 8600,
Switzerland}

\author{Shin-ichi Orimo}
\affiliation{
Institute for Materials Research, Tohoku University,
Sendai 980-8577, Japan}

\date{October 31, 2006}

\begin{abstract}
The thermodynamical stability of ${\rm Al(BH_4)_3}$ has been 
investigated using first-principles calculations based on density 
functional theory. The heats of formation are obtained to be $-132$ 
and $-131~{\rm kJ/mol}$ without the zero-point energy corrections for 
$\alpha$- and $\beta$-${\rm Al(BH_4)_3}$, respectively, which are 
made up of discrete molecular ${\rm Al(BH_4)_3}$ units. It is 
predicted correctly that the $\alpha$ phase is more stable than the 
$\beta$ phase. The energy difference between the solid phases and the 
isolated molecule is only about 10~kJ/mol. An analysis of the 
electronic structure also suggests the weak interaction between ${\rm 
Al(BH_4)_3}$ molecules in the solid phases. It is confirmed that 
${\rm Al(BH_4)_3}$ obeys the linear relationship between the heat of 
formation and the Pauling electronegativity of the cation, which has 
been proposed in our previous study 
[Nakamori {\it et al.}, Phys. Rev. B {\bf 74}, 045126 (2006)].
\end{abstract}
\pacs{65.40.-b, 71.20.Ps, 61.66.Fn}

\maketitle

%--------*---------*---------*---------*---------*---------*---------%

\section{Introduction}
\label{sec:intro}

Metal borohydrides are potential candidates for hydrogen storage 
materials because of their high gravimetric density of hydrogen. 
Among borohydrides, alkali borohydrides such as ${\rm LiBH_4}$ and 
${\rm NaBH_4}$ are well known and their properties have been 
investigated moderately.\cite{B1,B2,B3,B4,LiBH4,IMD,IMD2} Since 
these compounds are thermodynamically too stable and desorb hydrogen 
only at elevated temperatures, it is required to decrease their 
hydrogen desorption temperatures for practical applications. For 
${\rm LiBH_4}$, the reduction of the enthalpy change for the hydrogen 
desorption reaction has been attained by mixing with 
additives,\cite{Aoki,Pinkerton,Vajo} whose desorption temperatures 
are lower than that of pure ${\rm LiBH_4}$ by a few hundreds kelvins. 

Although borohydrides composed of non-alkali metals can be found in 
literature,\cite{MBH4} a little is known for their properties. In 
this 
context, we have recently investigated the thermodynamical stability 
of several metal borohydrides, $M{\rm (BH_4)}_n$~($M=$~Li, Na, K, Cu, 
Mg, Zn, Sc, Zr, and Hf), both theoretically and 
experimentally.\cite{HvsEN} It has been found that the stability of 
metal borohydrides shows a good correlation with the 
electronegativity of cations $M$. We have also determined the 
structural parameters experimentally for ${\rm Ca(BH_4)_2}$ and its 
fundamental properties have been predicted 
theoretically.\cite{CaB2H8}

Aluminum borohydirde ${\rm Al(BH_4)_3}$ is liquid at ambient 
temperatures with the melting point of 209~K. The structures of the 
solid phase have been investigated by x-ray diffraction 
measurements.\cite{Aldridge} Cooling liquid ${\rm Al(BH_4)_3}$, the 
orthorhombic $\beta$ phase was initially grown and then the 
transition to the monoclinic $\alpha$ phase occurred at a 
temperatures in the range 180-195~K.
In this study, we predict the thermodynamical stability of ${\rm 
Al(BH_4)_3}$ and reexamine the correlation between the stability of 
borohydrides and the cation electronegativity. 

\section{Method}
\label{sec:method}

The present calculations have been performed using the ultrasoft 
pseudopotential method\cite{USP} based on density functional 
theory.\cite{DFT} The generalized gradient approximation\cite{PBE}
(GGA) is adopted for the exchange-correlation energy. 
The cutoff energies used in this study are 15 and 120
hartrees for the pseudowave functions and the charge density, 
respectively. The $k$-point grids for the Brillouin zone integration
are generated so as to make the edge lengths of the grid elements as
close to the target value of 0.08~${\rm bohr^{-1}}$ as possible. 
These computational conditions give good convergence for the total 
energy within 1~meV/atom.
The computational details can be found in Ref.~\onlinecite{LiBH4} 
and the references therein.

\section{Results and discussion}

\subsection{Thermodynamical stability of ${\rm Al(BH_4)_3}$}
\label{sec:resula}

The heat of formation for the following reaction is considered, that 
is, the formation of ${\rm Al(BH_4)_3}$ from the elements:
\begin{equation}
\label{eq:reaction}
{\rm Al} + {\rm 3B} + {\rm 6H_2} \rightarrow {\rm Al(BH_4)_3}.
\end{equation}
The heat of formation is estimated from the difference of the total 
energies between the left- and right-hand sides of 
Eq.~(\ref{eq:reaction}).
In this section, we ignore the zero-point energy (ZPE) 
contribution to the heat of formation.

\begin{figure}
\begin{center}
\includegraphics[width=8cm]{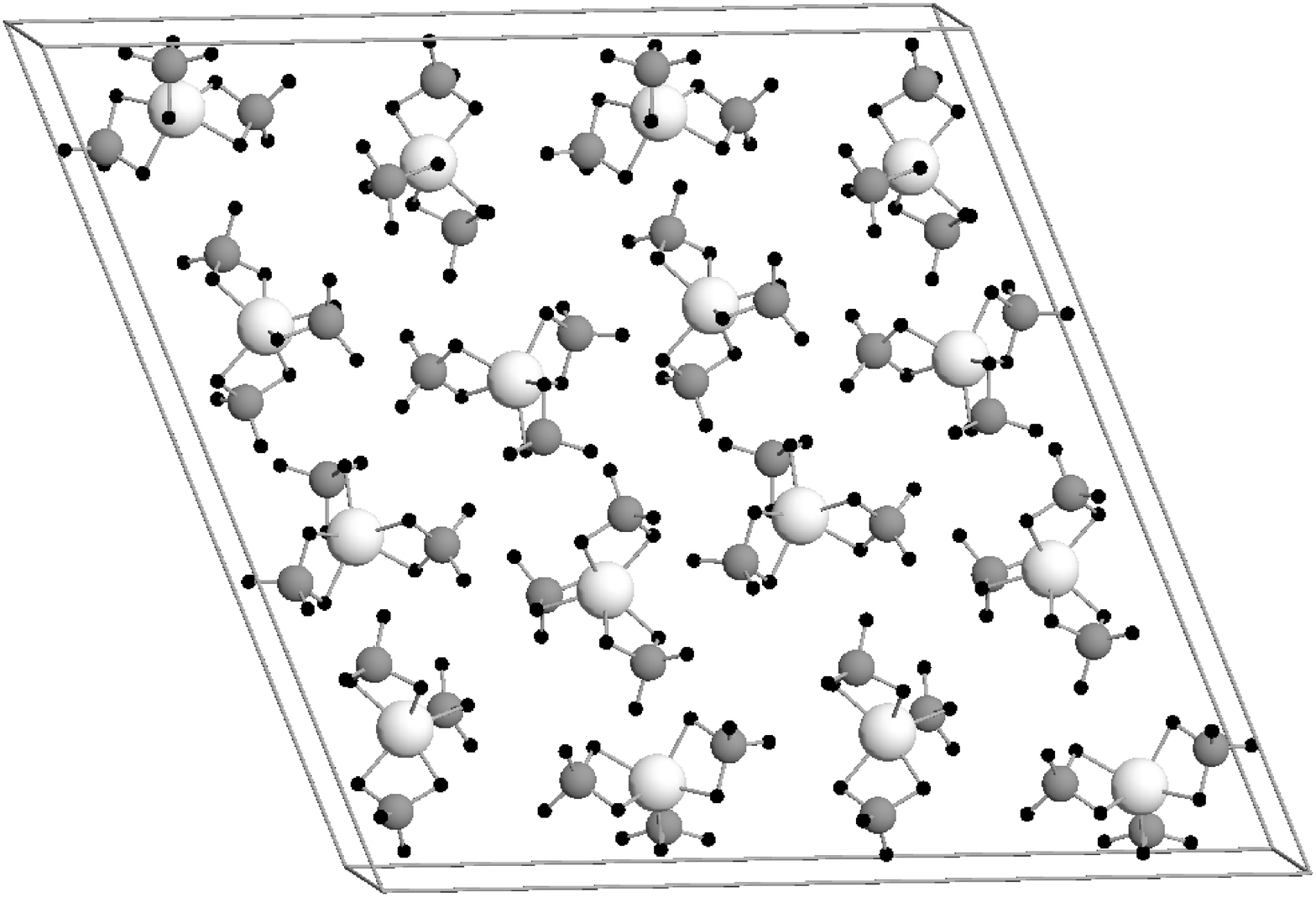}\\
(a) $\alpha$-${\rm Al(BH_4)_3}$ \\[10mm]
\includegraphics[width=8cm]{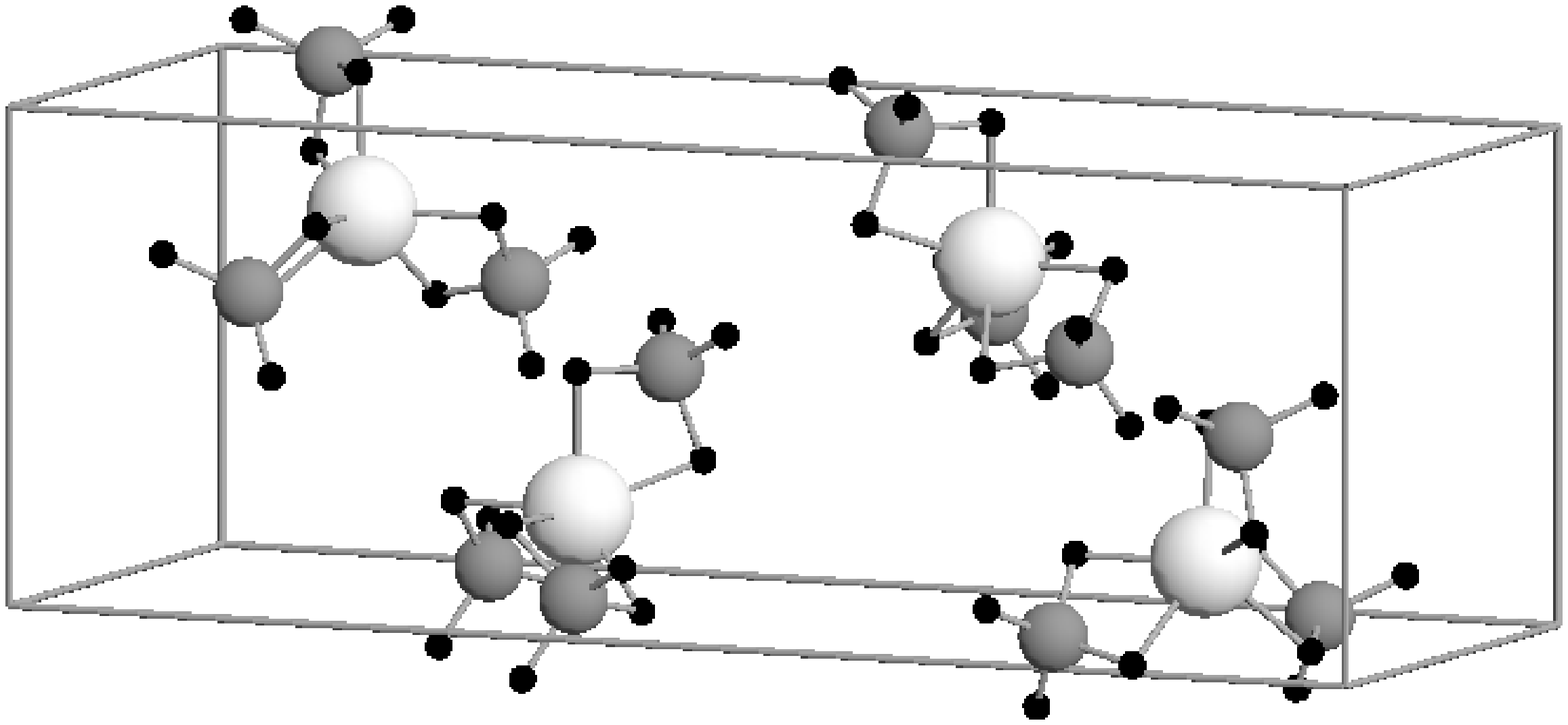}\\
(b) $\beta$-${\rm Al(BH_4)_3}$ \\
\end{center}
\caption{
Crystal Structure of (a) $\alpha$- and (b) $\beta$-${\rm 
Al(BH_4)_3}$. White, gray, and black spheres indicate Al, B, and H 
atoms, respectively.
\label{fig:str}}
\end{figure}

The crystal structures of $\alpha$- and $\beta$-${\rm Al(BH_4)_3}$ 
are shown in Figs.~\ref{fig:str}~(a) and (b), respectively.
Both phases 
are made up of discrete molecular ${\rm Al(BH_4)_3}$ units in which 
three boron atoms form a planar ${\rm AlB_3}$ skeleton and are bonded to 
Al atom by two bridging hydrogen atoms. The packing of the molecules 
is slightly denser in the $\alpha$ phase than in the $\beta$ phase. 
The 
structural optimization is performed for both phases starting from 
the experimental configurations,\cite{CCDC1,CCDC2} where the atomic
positions as well as the lattice vectors are fully relaxed. 

The optimized structural parameters are given in 
Tables~\ref{table:stra} and \ref{table:strb} for $\alpha$- and 
$\beta$-${\rm Al(BH_4)_3}$, respectively.
\begin{table*}
\caption{\label{table:stra}
Optimized structural parameters for $\alpha$-${\rm Al(BH_4)_3}$.
Space group: $C2/c$ (No.~15). $a=22.834$, $b=6.176$ and 
$c=22.423$~{\AA}, and $\beta=111.67^\circ$.
The unit cell contains two independent ${\rm Al(BH_4)_3}$ molecules.
The experimental lattice parameters are $a=21.917$, $b=5.986$ and 
$c=21.787$~{\AA}, and $\beta=111.90^\circ$.\cite{Aldridge,CCDC1}
}
\begin{ruledtabular}
\begin{tabular}{lccccclcccc}
  & Wyckoff & \multicolumn{3}{c}{coordinates} & &
  & Wyckoff & \multicolumn{3}{c}{coordinates} \\
\cline{3-5} \cline{9-11}
  & position & $x$ & $y$ & $z$ & &
  & position & $x$ & $y$ & $z$ \\
\hline
 Al     & 8f &   0.3797 &   0.5943 &   0.8366 & &
 Al     & 8f &   0.3659 &   0.1127 &   0.5947 \\
 B1     & 8f &   0.3205 &   0.3121 &   0.8239 & &
 B1     & 8f &   0.3626 &  $-0.1633$ &   0.5349 \\
 H1$_a$ & 8f &   0.3809 &   0.3078 &   0.8439 & &
 H1$_a$ & 8f &   0.3860 &  $-0.1645$ &   0.5965 \\
 H1$_b$ & 8f &   0.3002 &   0.5071 &   0.8128 & &
 H1$_b$ & 8f &   0.3437 &   0.0276 &   0.5138 \\
 H1$_c$ & 8f &   0.3067 &   0.2479 &   0.8677 & &
 H1$_c$ & 8f &   0.3163 &  $-0.2759$ &   0.5191 \\
 H1$_d$ & 8f &   0.3019 &   0.2214 &   0.7725 & &
 H1$_d$ & 8f &   0.4058 &  $-0.2032$ &   0.5180 \\
 B2     & 8f &   0.3899 &   0.7542 &   0.7555 & &
 B2     & 8f &   0.2793 &   0.2182 &   0.6028 \\
 H2$_a$ & 8f &   0.4165 &   0.5751 &   0.7794 & &
 H2$_a$ & 8f &   0.3147 &   0.0676 &   0.6368 \\
 H2$_b$ & 8f &   0.3516 &   0.8141 &   0.7820 & &
 H2$_b$ & 8f &   0.3067 &   0.3176 &   0.5703 \\
 H2$_c$ & 8f &   0.3575 &   0.7139 &   0.7005 & &
 H2$_c$ & 8f &   0.2749 &   0.3494 &   0.6407 \\
 H2$_d$ & 8f &   0.4309 &   0.8894 &   0.7685 & &
 H2$_d$ & 8f &   0.2332 &   0.1306 &   0.5652 \\
 B3     & 8f &   0.4298 &   0.7324 &   0.9297 & &
 B3     & 8f &   0.4542 &   0.2773 &   0.6427 \\
 H3$_a$ & 8f &   0.3703 &   0.7520 &   0.8985 & &
 H3$_a$ & 8f &   0.4187 &   0.3056 &   0.5838 \\
 H3$_b$ & 8f &   0.4541 &   0.6038 &   0.9002 & &
 H3$_b$ & 8f &   0.4282 &   0.1495 &   0.6705 \\
 H3$_c$ & 8f &   0.4349 &   0.6401 &   0.9787 & &
 H3$_c$ & 8f &   0.4573 &   0.4508 &   0.6683 \\
 H3$_d$ & 8f &   0.4531 &   0.9080 &   0.9304 & &
 H3$_d$ & 8f &   0.5010 &   0.1867 &   0.6426 \\
\end{tabular}
\end{ruledtabular}
\end{table*}
\begin{table}
\caption{\label{table:strb}
Optimized structural parameters for $\beta$-${\rm Al(BH_4)_3}$.
Space group: $Pna2_1$ (No.~33). $a=18.649$, $b=6.488$ and 
$c=6.389$~{\AA}.
The unit cell contains one independent ${\rm Al(BH_4)_3}$ molecule.
The experimental lattice constants are $a=18.021$, $b=6.138$ and 
$c=6.199$~{\AA}.\cite{Aldridge,CCDC2}
}
\begin{ruledtabular}
\begin{tabular}{lcccc}
  & Wyckoff & \multicolumn{3}{c}{coordinates} \\
\cline{3-5}
  & position & $x$ & $y$ & $z$ \\
\hline
 Al     & 4a & 0.8703 &   0.1558 &   0.2098 \\
 B1     & 4a & 0.7800 &   0.0057 &   0.0633 \\
 H1$_a$ & 4a & 0.8456 &  $-0.0341$ &   0.0213 \\
 H1$_b$ & 4a & 0.7751 &   0.1384 &   0.2112 \\
 H1$_c$ & 4a & 0.7552 &  $-0.1515$ &   0.1331 \\
 H1$_d$ & 4a & 0.7551 &   0.0858 &  $-0.0906$ \\
 B2     & 4a & 0.9168 &   0.0183 &   0.4855 \\
 H2$_a$ & 4a & 0.9353 &  $-0.0248$ &   0.2979 \\
 H2$_b$ & 4a & 0.8712 &   0.1661 &   0.4885 \\
 H2$_c$ & 4a & 0.9700 &   0.0865 &   0.5674 \\
 H2$_d$ & 4a & 0.8870 &  $-0.1319$ &   0.5547 \\
 B3     & 4a & 0.9115 &   0.4349 &   0.0722 \\
 H3$_a$ & 4a & 0.8623 &   0.4281 &   0.2121 \\
 H3$_b$ & 4a & 0.9337 &   0.2529 &   0.0277 \\
 H3$_c$ & 4a & 0.9619 &   0.5214 &   0.1520 \\
 H3$_d$ & 4a & 0.8837 &   0.5009 &  $-0.0844$ \\
\end{tabular}
\end{ruledtabular}
\end{table}
The lattice constants are 
overestimated: The maximum deviation from the experimental values is 
4~\% for the $a$ axis in the $\alpha$ phase and 6~\% for the $b$ axis 
in the $\beta$ phase, which are larger than the errors of typical GGA 
calculations. The heats of formation are obtained as $\Delta 
H=-132~{\rm kJ/mol}$ for $\alpha$-${\rm Al(BH_4)_3}$ and $\Delta 
H=-131~{\rm kJ/mol}$ for $\beta$-${\rm Al(BH_4)_3}$. It is predicted 
correctly that the $\alpha$ phase is more stable than the $\beta$ 
phase.

The relatively large deviations found for the lattice constants are 
probably due to the weak interaction between molecular ${\rm 
Al(BH_4)_3}$ units in the solid phases. The ${\rm Al(BH_4)_3}$ 
molecules are expected to be neutral and not bonded strongly to each 
other. In this case, van der Walls interactions play an important 
role, which can not be described properly by the current GGA 
functional. To check this problem, we repeat the structural 
optimization, where the lattice constants are kept fixed at the 
experimental values and the atomic positions are only relaxed. The 
obtained heats of formation are $-128$ and $-127~{\rm kJ/mol}$ for 
the $\alpha$ and $\beta$ phases, respectively. It can be confirmed 
that the relaxations of the lattice vectors give only a minor effect 
for the energetics of ${\rm Al(BH_4)_3}$.

For comparison, the calculation is performed on molecular ${\rm 
Al(BH_4)_3}$ which has $D_3$ symmetry. A hexagonal supercell with 
lattice constants of $a=c=13.2$~{\AA} is used with single 
$\Gamma$-point sampling. The heat of formation is predicted to be 
$\Delta H=-122~{\rm kJ/mol}$. As expected, $\Delta H$ of molecular 
${\rm Al(BH_4)_3}$ is slightly higher than those of the solid states 
by about $10~{\rm kJ/mol}$. This small energy difference must be 
closely related to low melting point of ${\rm Al(BH_4)_3}$. In 
Talbe~\ref{table:bond}, we compare the bond lengths and angles 
obtained for molecular ${\rm Al(BH_4)_3}$ with those for two solid 
phases. 
\begin{table*}
\caption{\label{table:bond}
Comparison of the bond lengths $d$~({\AA}) and angles $\theta$~($^\circ$) obtained 
for molecular ${\rm Al(BH_4)_3}$ and two solid phases. For $d$(Al-H) 
and $\theta$(H-Al-H), H atoms forming the bridge bond are 
only cosidered.
}
\begin{ruledtabular}
\begin{tabular}{lcccccc}
 & \multicolumn{3}{c}{Calc.} & & \multicolumn{2}{c}{Expt.\tablenotemark[1]} \\
 \cline{2-4} \cline{6-7}
 & molecule & $\alpha$ phase & $\beta$ phase & & $\alpha$ phase &
$\beta$ phase \\
\hline
d(Al-B)   & 2.16      & 2.15-2.16 & 2.15-2.16 & & 2.10-2.14 & 2.10-2.13 \\
d(Al-H)   & 1.77      & 1.77-1.78 & 1.77-1.78 & & 1.73-1.76 & 1.68-1.75 \\
d(B-H)    & 1.20-1.29 & 1.20-1.29 & 1.21-1.28 & & 0.99-1.14 & 0.99-1.14 \\ 
$\theta$(B-Al-B) & 120       & 117-122   & 119-122   & & 119-121   & 119-123 \\
$\theta$(H-Al-H) & 73        & 73        & 73        & & 65        & 63-65   \\
$\theta$(H-B-H)  & 106-122   & 105-121   & 105-121   & & 103-121   & 106-130 \\
\end{tabular}
\tablenotetext[1]{Reference \ \onlinecite{Aldridge}.}
\end{ruledtabular}
\end{table*}
The geometries of ${\rm Al(BH_4)_3}$ molecules are 
essentially unchanged when forming the solid phases. 
The agreement with the
experimental data is good except for some of hydrogen related
parameters. This is probably caused by the experimental difficulty in
identifying H positions due to their weak x-ray scattering power.

\begin{figure}
\begin{center}
\includegraphics[width=8cm]{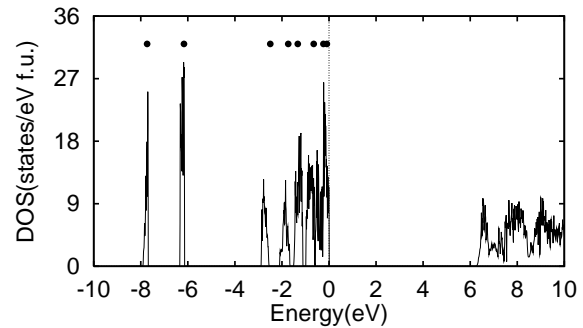}
\end{center}
\caption{
Electronic density of states for $\alpha$-${\rm Al(BH_4)_3}$. The 
origin of energy is set to be the top of valence states. The energy 
positions of the occupied states for molecular ${\rm Al(BH_4)_3}$ are 
indicated by solid circles for comparison purpose.
\label{fig:dos}}
\end{figure}

Figure~\ref{fig:dos} depicts the electronic density of states for 
$\alpha$-${\rm Al(BH_4)_3}$. 
The electronic structure is nonmetallic 
with a calculated gap of 6.0~eV. The occupied states consist of 
several sharp peaks whose energy positions correspond well to those 
of molecular ${\rm Al(BH_4)_3}$. The similar correspondence can be 
found for $\beta$-${\rm Al(BH_4)_3}$.
These also suggest the weak interaction 
between ${\rm Al(BH_4)_3}$ molecules in the solid phases.

\subsection{Correlation between stability of borohydrides and cation 
electronegativity}
\label{sec:resulb}

In Ref.~\onlinecite{HvsEN}, we have found a good correlation between the 
heat of formation $\Delta H$ and the Pauling electronegativity of the 
cation $\chi_P$, when $\Delta H$ is normalized by the number of ${\rm 
BH_4}$ complexes in the formula unit. This correlation can be 
represented by the linear 
relationship, $\Delta H=248.7\chi_P-390.8$ in the unit of ${\rm 
kJ/mol~BH_4}$. In this section, we refine this relation using the 
additional results for $\alpha$-${\rm Al(BH_4)_3}$ and ${\rm 
Ca(BH_4)_2}$.\cite{CaB2H8}

\begin{figure}
\begin{center}
\includegraphics[width=8cm]{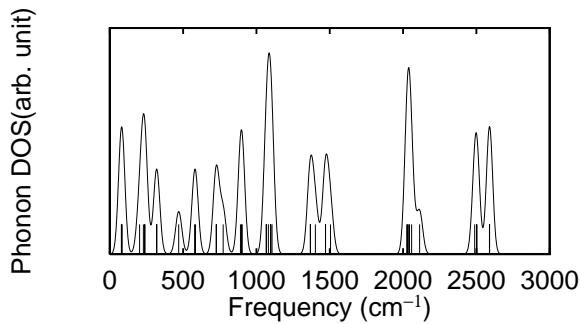}
\end{center}
\caption{
Normal eigenmode frequencies for molecular ${\rm Al(BH_4)_3}$. The 
calculated frequencies are indicated by vertical bars and the 
Gaussian broadening with a width of $30~{\rm cm}^{-1}$ is used.
\label{fig:gdos}}
\end{figure}

In order to estimate the ZPE contribution to $\Delta H$ for ${\rm 
Al(BH_4)_3}$, the normal eigenmode frequencies are calculated for 
molecular ${\rm Al(BH_4)_3}$. The weak interaction between ${\rm 
Al(BH_4)_3}$ molecules in the solid phases justifies this 
treatment. The result is shown in Fig.~\ref{fig:gdos}, where the 
obtained frequencies agree fairly well with the results of 
the experiment\cite{Coe} and the quantum 
chemistry calculation.\cite{Jensen}
The ZPE contribution to $\Delta 
H$ is predicted to be $36~{\rm kJ/mol~BH_4}$. This value is quite 
close to the approximated value of $33~{\rm kJ/mol~BH_4}$ adopted in 
Ref.~\onlinecite{HvsEN}, which has been obtained using the molecular 
approximation for a ${\rm [BH_4]^-}$ anion.
To keep the consistency with the previous treatment, we decide 
to use the approximated value of $33~{\rm kJ/mol~BH_4}$, where the
normalized heats of formation with the ZPE correction become
$-11$ and $-155~{\rm kJ/mol~BH_4}$ for ${\rm Al(BH_4)_3}$ and 
${\rm Ca(BH_4)_2}$, respectively.

\begin{figure}
\begin{center}
\includegraphics[width=8cm]{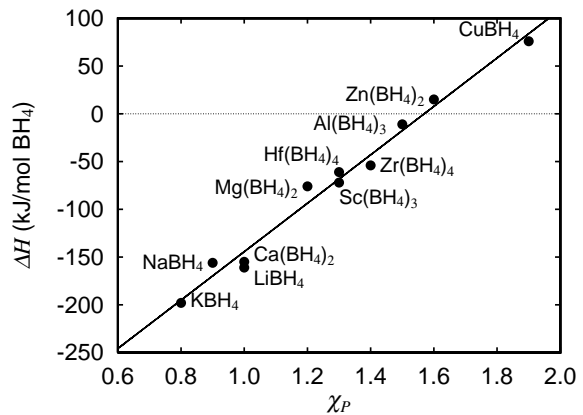}
\end{center}
\caption{
Relation between the heat of formation $\Delta H$ and 
the Pauling electronegativity of the cation, $\chi_P$.  The straight 
line indicates the result of the least square fitting, $\Delta 
H=253.6\chi_P -398.0$. The zero-point energy 
contributions to $\Delta H$ are approximately taken into 
consideration (see the text). 
\label{fig:HvsEN}}
\end{figure}

In Fig.~\ref{fig:HvsEN}, the normalized heat of formation $\Delta H$ 
with the ZPE correction
as a function of the Pauling electronegativity of the cation $\chi_P$ 
is plotted.\cite{MgB2H8}
It can be found that the results for ${\rm Al(BH_4)_3}$ 
and ${\rm Ca(BH_4)_2}$ also obey the linear relationship.
The least square fitting yields
\begin{equation}
\Delta H=253.6\chi_P -398.0,
\end{equation}
with an absolute mean error of $9.6~{\rm kJ/mol~BH_4}$. The addition 
of two new data points causes only a little change in the 
coefficients 
for the linear relation. The present study provides the additional 
support for the linear correlation between $\Delta H$ and $\chi_P$. The 
Pauling electronegativity of the cation $\chi_P$ is a good indicator 
to estimate the thermodynamical stability of borohydrides.

\section{Conclusions}
\label{sec:conc}

In this study, we have predicted the thermodynamical stability of 
${\rm Al(BH_4)_3}$ and reexamined the correlation between the 
stability of borohydrides and the cation electronegativity. 

The heats of formation are obtained as $-132$ and $-131~{\rm kJ/mol}$ 
without the zero-point energy corrections for $\alpha$- and 
$\beta$-${\rm Al(BH_4)_3}$, respectively, which are made up of 
discrete molecular ${\rm Al(BH_4)_3}$ units. It is correctly 
predicted that the $\alpha$ phase is more stable than the $\beta$ 
phase. The energy difference between the solid phases and the 
isolated molecule is small and the interaction between the molecules 
in the solid phases is expected to be weak. This is most likely 
related to low melting point of ${\rm Al(BH_4)_3}$.

The linear correlation between the heat of formation and the Pauling 
electronegativity of the cation, which has been proposed in the 
previous study,\cite{HvsEN} is also held for ${\rm Al(BH_4)_3}$ as 
well as for ${\rm Ca(BH_4)_2}$. The Pauling electronegativity of the 
cation is a good indicator to estimate the thermodynamical 
stability of borohydrides.

\begin{acknowledgments}

The authors would like to thank M. Aoki, T. Noritake, M. Matsumoto, 
H. -W. Li and S. Hyodo for valuable discussions. This study is partially 
supported by the New Energy and Industrial Technology Development 
Organization (NEDO), International Joint Research under the 
^^ ^^ Development for Safe Utilization and Infrastructure of hydrogen" 
Project (2005-2006).

\end{acknowledgments}

%\clearpage

\end{document}